# Ultrafast Pulse Radiolysis Using a Terawatt Laser Wakefield Accelerator

Dmitri A. Oulianov, Robert A. Crowell[*], David J. Gosztola, Ilya A. Shkrob, Oleg J. Korovyanko, and Roberto C. Rey-de-Castro
*Chemistry Division, Argonne National Laboratory, 9700 South Cass Avenue, Argonne, IL 60439*

**Abstract**

We report the first ultrafast pulse radiolysis transient absorption spectroscopy measurements from the Terawatt Ultrafast High Field Facility (TUHFF) at Argonne National Laboratory. TUHFF houses a 20 TW Ti:sapphire laser system that generates 2.5 nC sub-picosecond pulses of multi-MeV electrons at 10 Hz using laser wakefield acceleration. The system has been specifically optimized for kinetic measurements in a pump-probe fashion. This requires averaging over many shots which necessitates stable, reliable generation of electron pulses.  The latter were used to generate excess electrons in pulse radiolysis of liquid water and concentrated solutions of perchloric acid. The hydronium ions in the acidic solutions react with the hydrated electrons resulting in the rapid decay of the transient absorbance at 800 nm on the picosecond time scale. Time resolution of a few picoseconds has been demonstrated.  The current time resolution is determined primarily by the physical dimensions of the sample and the detection sensitivity.  Subpicosecond time resolution can be achieved by using thinner samples, more sensitive detection techniques and improved electron beam quality.

___________________________________________________

[*] To whom correspondence should be addressed: *Tel* 630-2528089, *FAX* 630-2529570, *e-mail:* rob_crowell@anl.gov.

## I. INTRODUCTION

One of the most common yet still poorly understood sources for chemical activation energy in the Universe is ionizing radiation. Radiation chemists study the rapid, energetic reactions that are initiated by the interaction of the ionizing radiation, such as high energy electrons, with matter. Understanding of these reactions impacts many fields including the design of nuclear reactors, radioactive waste management, radiation therapy, polymer processing, and planetary- and astro- physics.[1]

As highly ionizing radiation passes through the condensed phase, energy is occasionally deposited along the trajectory which is commonly referred to as the radiolytic track. The points at which the relativistic electrons interact with condensed matter result in the formation of radiolytic spurs, viz. tight clusters of excited and ionized molecules. For 1-10 MeV electrons in water, the spurs are 5-10 nm in diameter and are separated by 200-400 nm along the track.[2] The rapid physical and chemical events that take place in the radiolytic spur occur on the picosecond and shorter time scales. Detailed knowledge of fundamental spur processes, such as thermalization, solvation, and the reactions of short-lived energetic species (excited states and ions), is critical in order to develop an understanding of the mechanism(s) for radiation damage.[1,3,4] Quite often these intraspur reactions involve exotic excited states and cross-reactions between ionized/excited molecules.[5-7] Generation of such otherwise hard-to-access states, the unusual reaction regimes, and the inherently non-uniform deposition of the primary ionization/excitation events make it difficult if not impossible to simulate rapid radiolytic processes using ultrafast laser sources.[8-14] Studies of the fundamental processes of radiation induced reactions require an ultrafast source of ionizing radiation. Furthermore, this source must be accurately synchronized to an ultrafast laser that is used as a probe source for the detection of the short-lived species and their kinetics.

For the past thirty years the most common source of ionizing radiation used for subnanosecond pulse radiolysis studies has been radio frequency (RF) linear electron accelerators (linacs) based on thermionic guns.[15-19] The time resolution of transient absorption pulse radiolysis experiments using this type of linac has, for the most part, been limited to 20-30 ps at 10-100 Hz.[20] Time resolution of about 5 ps has been realized for emission measurements using a streak camera for detection.[21] Recently, new types of electron accelerators based on laser-driven photocathode linacs have been specifically developed for radiation chemistry.[22-27] At Osaka University scientists have produced single electron bunches that are less than 100 fs in duration (at 32 MeV) using a photocathode linac.[24,28] Despite the ultrashort bunch lengths produced with this new generation of accelerators the best time resolution attainable for transient absorption pulse radiolysis experiments is typically 4-10 ps.[26,29-31] This is primarily due to detection sensitivity and timing jitter between the laser and accelerator. However, the improved time resolution that photocathode based accelerators offer over thermionic linacs has opened up the door for new areas of chemical research. Recent picosecond pulse radiolysis experiments that take advantage of the new time window include a re-evaluation of the initial radiolytic yield of the hydrated electron in water,[30] ultrafast studies of quantum confinement in semiconducting scintillators,[32] fast charge transfer mechanisms,[33] radiolytic reactions[31,34,35] and radiolysis of supercritical liquids.[36,37]

For photocathode linacs, the experimental time resolution is highly dependent upon the degree of synchronization between the phase of the RF field and the laser pulse that generates the photoelectrons at the photocathode as well as the synchronization between the resulting electron pulse and the probe laser pulse. For example, to generate an 800 fs electron pulse, the time jitter between the RF and the laser must be reduced to better than 500 fs.[38] This level of stability is difficult to reliably achieve in practice and is one of the factors limiting the usefulness of this approach for ultrafast pulse radiolysis studies.

At Argonne National Laboratory, a different approach that capitalizes on recent advances in laser wakefield acceleration has been pursued. Since the advent of chirped pulse amplification[39] significant advances have been made in ultrafast high power femtosecond laser technology. Ti:Sapphire based table-top terawatt ($T^3$) laser systems are now routinely capable of generating peak powers in excess of $10^{13}$ W (that is equivalent to 500 mJ delivered in 50 fs).[40] By focusing terawatt laser pulses to irradiances exceeding $10^{18}$ W/cm$^2$ in a pulsed supersonic helium gas jet, it is possible to generate subpicosecond electron pulses with a charge of a few nanoCoulombs and to accelerate these electrons to energies in the MeV range.[41]

While this technology has been developed primarily by physicists interested in studying laser-plasma interactions and acceleration physics, nonlinear quantum electrodynamics, and the development of next generation x-ray light source,[42] it should be also suitable for subpicosecond pulse radiolysis experiments, provided that continued operation at a reasonable high repetition rate (conducive for pump-probe kinetic measurements) and sufficiently large electron flux (ca. 0.5 nC/mm$^2$) are possible. In most of the physics experiments these objectives were not the goals (the focus being, for example, monochromaticity of the electron energy, more efficient electron and proton acceleration, etc.) and most of the experiments were carried out in what is, essentially, single pulse mode with poor reproducibility of the electron pulse characteristics. By contrast, for chemical applications, a typical pump-probe measurement requires averaging of $10^3$-$10^4$ shots. In this case it is critical to have a source with reproducible parameters that operates at a reasonable repetition rate (maximum of 10 Hz for a typical laser wakefield accelerator). Chemical applications present new demands to the laser wakefield method itself.

It has been shown that laser wakefield accelerators can produce sufficient charge per pulse for detection of radiolysis products using transient absorbance (TA) laser spectroscopy. Using the laser wakefield accelerator (LWA) at the University of Michigan (Center for Ultrafast Optical Studies), subpicosecond electron pulses were generated by focusing terawatt laser pulses into a supersonic helium jet and subsequently used to ionize liquid water. The ionization of water results in the generation of metastable hydrated electrons ($e^-_{aq}$) in which the *s*-function of the excess electron occupies a solvation cavity of ca. 0.4 nm in diameter that is formed by dangling OH groups of 6-8 water molecules.[43] This species plays the central role in water radiolysis and it strongly absorbs in the visible and near infrared, exhibiting the extinction coefficient of ca. $2\times10^4$ M$^{-1}$ cm$^{-1}$ at the absorption maximum of 720 nm.[44] Using the Michigan LWA, hydrated electron concentrations as high as $2\times10^{-5}$ mol dm$^{-3}$ ($1.2\times10^{16}$ cm$^{-3}$) were generated and the attachment of $e^-_{aq}$ to traces of O$_2$ in water was followed on the nano-to-microsecond timescale.[45] More recently the Malka and co-workers[46] have demonstrated approximately 20 ps resolution using a LWA for pulse radiolysis of liquid water. Below we describe the next generation setup that is capable of producing 10 Hz train of 2.5 nC subpicosecond pulses for hours at a time, which enabled us to perform pulse-probe measurements of chemical kinetics on a picosecond time scale using Argonne's Terawatt Ultrafast High Field Facility (TUHFF) LWA.

In a LWA, the electron and laser pulses are inherently synchronized so the time jitter issues associated with photocathode linacs are not an issue. The ultimate time resolution should depend only upon the cross correlation between the laser and electron pulses and the physics of the electron beam interaction with the sample. However, there are three main constraints that limit the current time-resolution to about 2 ps. One of the major constraints that limits the ultimate time resolution for ultrafast radiolysis measurements, in general, is the fact that relativistic electrons move through the sample at velocities close to the speed of light, $c$, whereas the probe photons travel at a substantially lower velocity, $c/n$ (where $n$ is the refractive index of the medium). Another fundamental constraint is that, unlike the linacs, the present laser wakefield accelerators do not reliably produce monoenergetic beams (although recent work indicates that this will change in the near future).[47-57] The TUHFF LWA has an electron spectrum that is Maxwellian, with the median energy corresponding to that of the electron plasma in the jet. The energy spread is more than 100% realtive to the mean. This dispersion in energy results in the dispersion of arrival times at the sample as the electron pulse travels through space; furthermore, since electrons of different energies are stopped differently by the sample, via scattering and loss of energy to the medium, there is additional temporal spread of the electron pulse in the sample. Therefore, to a much larger degree than in laser TA experiments, time resolution is limited by the nature and the physical dimensions of the sample, and the dispersion of the electron beam. Improvements in time resolution will require thin samples and the development of more sensitive detection techniques.

The linear energy transfer for a typical aqueous sample is 2-5 MeV/cm (for 1-100 MeV electrons)[58] and the typical radiation yield ($G$-value) of the species of interest is less than 5 per 100 eV of absorbed energy.[58,59] This means that in order to produce a detectable TA signal (>$10^{-4}$) from the radiation-induced species one has to have a sufficiently thick sample (a few millimeters) to stop the electrons. This is not the case for low energy photons and electrons where very thin samples can absorb most of the energy. Therefore, ultrafast pulse radiolysis using fast electrons (of the range that is important for applications) is always a compromise between time resolution and sensitivity of detection. In our present experimental configuration, the time resolution is limited not by the characteristics of the LWA source or synchronization issues, but by the sensitivity of the TA measurement itself and the physics of the electron interaction. For other high energy electron sources (e.g., photocathode based linacs), that is not the case. Below, we describe the laser wakefield accelerator that provides subpicosecond electron pulses for chemical physics research and demonstrate its use for ultrafast studies of aqueous pulse radiolysis.

## II. THE LASER SYSTEM.

The 20 TW laser facility[3] consists of a three-stage chirped-pulse amplified Ti:Sapphire laser system running at 10 Hz. After stretching the 15 fs fwhm seed pulse train from an oscillator to 440 ps in a double-pass single grating stretcher, a Pockels cell pulse-picker (Medox) is used to lower the repetition rate to 10 Hz. The first amplifier has a multi-pass ring design which uses a 10 mm diameter, 10 mm long Brewster angle amplifier rod (all Ti:Sapphire crystals used were from Crystal Systems) that is pumped with 25 mJ of 532 nm light that is split off of the output of a frequency-doubled Nd:YAG laser (all Nd:YAG lasers were from Spectra Physics). After six passes through the crystal, the pulse passes through a second Pockels cell and polarizer to minimize the amplified spontaneous emission background. The pulse energy at this point is 2.5 mJ with a beam diameter of 0.8 cm ($1/e^2$). The second amplifier consists of a 4-pass bowtie arrangement around an anti-reflection coated 20 mm diameter, 20 mm long cylindrical Ti:sapphire crystal. This crystal is pumped from both sides with a total of 0.85 J at 532 nm. A -1.5 m lens is placed after the first pass to compensate for the thermal lensing that occurs in the amplifier rod. The output of the second amplifier is spatially filtered and expanded to 12 mm ($1/e^2$). At this point, the pulse energy is 330 mJ. A beam splitter is used to divert 60 mJ of this beam to a two-grating pulse compressor and is used as a probe for transient absorption spectroscopy. The remaining beam is further amplified in a cryogenically cooled multi-pass amplifier consisting of a 30 mm diameter, 30 mm long antireflection coated cylindrical Ti:sapphire crystal held at 77 K in a liquid nitrogen cryostat (Janis Cryogenics). This last crystal is pumped from both sides with a total energy of 2.4 J at 532 nm. After two passes, the total energy at 800 nm is 1.1 J. In order to minimize the amplification of slight aberrations in the beam profile, the spatial profile of the beam is rotated by 90° after the first pass. After the second pass, the beam is expanded to 50 mm ($1/e^2$) and is directed into a vacuum chamber housing a two-grating pulse compressor resulting in a 600 mJ, 35 fs fwhm pulse. A more detailed description of the laser system can be found elsewhere.[3]

Electron pulses of relativistic energies (1.5-20 MeV) are generated inside a target vacuum chamber by focusing the laser beam onto a 1.2 mm diameter supersonic helium jet (Figure 1) using a 50 cm focal length off-axis gold parabolic mirror (Janos Technology). A solenoid valve (Parker) fitted with a supersonic nozzle is pressurized to 70 bar and opens for 2 ms. The valve is mounted on a motorized three-axis stage which allows for the optimization of the electron yield by fine movement of the jet within the focus of the laser. The nozzle design and characterization were based upon the work of the Malka[60] and Umstadter[61] groups. The performance of the He jet has been assessed using laser interferometric imaging.[60] We have tested different types of nozzle designs and pressure conditions and selected the one which produced the highest electron charge (see below).

## III. CHARACTERIZATION OF THE ELECTRON BEAM.

The beam charge was determined using a home-built Faraday cup (FC) with a 1.3 ns response time. The electrons were stopped in a 45 mm diameter, 25 mm thick copper disk housed inside a grounded aluminum box; the output was terminated into a 50 Ω load. Before hitting the cup, the electrons passed through a 400 μm thick aluminum shield. Good shielding of the FC from the electromagnetic pulse generated by the plasma in the target chamber was important for accurate charge measurements to a few picoCoulombs. The shielding also protects the cup from $He^+$ ions generated in the chamber. The short-lived negative current spike from the FC was sampled and integrated on a fast digital oscilloscope (LeCroy model LT354). Alternatively, the current was integrated using a boxcar integrator (Stanford Research Systems model 245) with a 10 ns gate that was calibrated against the oscilloscope signal. The latter mode was used during the kinetic measurements in order to monitor charge fluctuations. These were normally distributed, with the typical dispersion of 15-30%. Under optimum conditions, the typical charge per pulse was 1.5-3 nC, with at least 30% of the pulses > 2 nC. Importantly, this is the total electron charge integrated across the entire beam. Due to the large beam divergence (ca. 6-20$^o$) and the electron scattering in the jet, the copper disk (see below) and the sample itself, only a fraction of the beam passes through the sample. The sample (typically 10 mm x 10 mm square cell made of 1 mm thick fused silica, filled with liquid) was placed in an aluminum holder with a 10 mm diameter aperture at the back; typically 17 mm away from the jet. A rotating copper disk 200 µm thick, 10 cm diameter was placed directly in front of the sample to block the laser light. The disk was rotated during the interval between the laser pulses with a synchronized stepper rotor (New Focus model 8341-UHV). We found that ablation from the copper disk was much lower than that from an aluminum disk. The charge throughput through the empty cell in the holder was ca. 25%; and the throughput for the same cell filled with water was 0.8% indicating that almost all electrons were stopped by the sample.

Figure 2a shows the beam profile at the sample obtained by placing a thin radioluminescent screen (Kodak LANEX film) attached to a 1 mm thick glass plate inserted at the position of the sample. The emitted light was imaged on a digital camera equipped with the appropriate glass filter and a 0$^o$ dielectric mirror (to exclude scattered 800 nm light). The beam profile, after averaging over many shots, was approximately Gaussian with semi-major *1/e* axes of 11 and 12 mm (ca. 27 mm away from the jet), i.e., the divergence of the beam was ca. 20$^o$. There is also a hint of a tighter 12$^o$ cone for some of the pulses (Figure 2b), but the position of this feature varied from shot to shot and it was smoothed over when averaged over many pulses.

The energy spectrum of the electrons was found to vary across the beam, with higher energy electrons occupying a tighter cone. This feature reveals multiple elastic

9.

scattering in the jet and the metal foil used to block the light. The electrons generated in the plasma undergo multiple scattering events as they emerge from the jet. This scattering is stronger for low energy electrons and it results in a steep angular distribution of electron energies. Thus, even if the initial energy spectrum of the electrons is Maxwellian, as suggested by many experiments,[41,42,62] the spectrum of these electrons after scattering in the jet and the metal foil is not, as the low-energy electrons have greater scattering angles. This results in an angular dependence of electron energy.

Since we are interested only in the energy of the electrons impinging on the sample, a spectrometer was built to determine the average energies of the electrons passing through a copper collimator. This spectrometer consisted of two NdFeB magnets sandwiched between two steel bars and mounted on a copper plate. A soft steel shield was placed in front of this plate to screen stray magnetic field. A phosphor screen was placed 65 mm behind the aperture and imaged on a camera; a thin Al foil before this screen blocked the 800 nm light. The inhomogeneity of the magnetic field in the vertical direction was < 1%. The magnetic field $B(x)$ along the path of the light (axis $x$) was mapped using a Hall probe magnetometer. For every initial energy $E$ of the electron moving in the forward direction through the aperture, the relativistic equations of motion in the field $B(x)$ were integrated and vertical position $y_s(E)$ at which the curved electron trajectory intersects with the phosphor screen determined. For the geometry described, the minimum detectable energy $E$ was 1.5 MeV (corresponding to $y_s \approx 90$ mm) and the maximum energy that can be resolved using a collimator with a 5 mm diameter aperture was 20 MeV. The aperture was placed 135 mm away from the jet and the image of dispersed electrons on a LANEX screen was collected and averaged for 100 laser shots, with and without the jet pulsed (in order to subtract the background). An example of the dispersed electron beam is shown in Figures 3a and 3b. For the 5 mm diameter aperture (a 2° cone; see Figure 3a), the electron distribution peaked at $y_s \approx 13$ mm which corresponds to 7.5 MeV. The high energy tail extends to 20-25 MeV and there is a low-energy wing, with roughly 50% of electrons at the minimum (cut-off) energy of 1.5 MeV as compared with the maximum at 5-10 MeV. For a larger aperture of 32 mm diameter (a 13.5° cone) the position of the maximum shifts to 4 MeV and the energy spectrum is more Maxwellian (Figure 3b). There is a clear "tail" extending towards the higher energies. The probability to find the electron at 20 MeV is roughly 30% of that at 4 MeV. There is also a gradual onset of the distribution at 1-2 MeV. Once more, this gradual onset is due to the scattering of low-energy electrons into a wider cone.

A stack of 0.5-3 mm thick Al plates with the polyester-based radiochromic film (GAFCHROMIC MD-55, 270 μm thick, with the stopping power of ca. 2 MeV/cm at 1-3 MeV) pressed between the plates was exposed to 20-50 electron pulses, and the film was scanned with resolution of 200 dpi. The film is about 5 times more sensitive at 675 nm than at 550 nm, i.e., by scanning the same film in red, green, and blue (with the relative sensitivity of 15.7:3.4:1) over-exposure can be avoided. At maximum sensitivity (675 nm) the optical density is ca. 0.03 per 1 J/kg of absorbed radiation. The typical dose profile across the beam behind a 19 mm diameter, 6.4 mm thick copper aperture (the radiochromic film was placed 27.9 mm away from the jet plane) is shown in Figure 2. The distribution of the dose in nearly uniform within the 22° cone; on top of this broad distribution there is a narrower cone of 6° that carries less than a few per cent of the dose (this tighter distribution is from higher-energy electrons). Due to the electron scattering in the plates, this feature rapidly fades as the electrons penetrate through the stack, the profile becomes nearly Gaussian. By integrating over these profiles one can obtain the transverse profile of dose deposition shown in Figure 4. Maxwell's distribution $p(E; x=0) = \langle E \rangle^{-1} \exp(-E/\langle E \rangle)$ of electron energies was assumed for the electrons incident on the plates, where $\langle E \rangle$ is the temperature in energy units. Using the known dependence of stopping power $S(E)$ for Al as a function of electron energy $E$ one obtains for the linear energy transfer (LET) dose:

$$dE/dx = -S(E), \qquad (1)$$

from which the mean energy loss as a function of thickness $x$ of the material and charge throughput can be estimated. Convoluting this profile with the initial Maxwell distribution, one obtains the power spectrum of the electrons at a given depth $x$. The calculated dose deposition profile can be compared with the experimental one, with the mean energy as the only adjustable parameter. This optimization gives $\langle E \rangle \approx 2.3 \pm 0.3$ MeV, which thereby is the temperature of the electron plasma in energy units. The predicted spectrum of the electrons 1 mm inside the aqueous sample (integrated over the entire beam) is shown in Figure 5(a).

Using this spectrum it is easy to estimate the dispersion of arrival times *t(E)* for the electrons at the normal plane containing the optical path of the probe (Figure 5(b)). Most of this dispersion originates from low-energy electrons that move significantly slower than the speed of light but are abundant at the onset of the thermal distribution of energies. Some of these electrons are stopped by the 200 μm copper shield and the glass cell walls so only electrons with energies > 1 MeV arrive at the sample, which is ca. 70% of the total charge. For high-energy electrons (above the median energy) the dispersion is only 1.5 ps, but for low energy electrons that pass through the sample, it is ca. 2.8 ps, so the overall spread is ca. 4.3 ps fwhm. Using a thicker copper shield would result in better shaping of the electron spectrum: e.g., a 600 μm copper shield yields a 3 ps fwhm dispersion of arrival times (with 1.3 MeV cutoff and 45% of the electrons stopped before the sample). These are conservative estimates: due to the electron scattering, the spectrum of the electrons in the cone arriving at the sample is already depleted of the low-energy electrons. A more realistic estimate of this dispersion under the conditions of our experiment was about 2 ps fwhm.

## IV. DETECTION SYSTEM

The detection system was used to determine time evolution of transient absorbance at 800 nm (that is, the temporal change in the optical density, $\Delta OD$) using pump-probe methodology. The 800 nm beam from the second amplifier was compressed using a standard two-grating compressor to < 30 fs fwhm and suitably delayed to temporally overlap the electron and probe pulses at the sample. In one of the delay arms, a 80 cm double pass motorized delay stage (Velmex) was inserted. The probe light was split 1:1, with one beam serving as a reference, to compensate for fluctuations in the probe light intensity. The second beam, used to probe the sample, was steered into the target chamber where it intersected with the electron beam at 90° inside the sample cell.

A pair of photodiode detectors (Si FND-100Q, -100 V bias) were encased in a Faraday cage and shielded behind a 10 cm thick lead wall erected 3 m away from the target chamber so that the electromagnetic interference from plasma discharge and stray gamma rays were mostly eliminated (to less than $10^{-6}$ OD). The probe beams were spatially and spectrally filtered so that very little light from the plasma spark reached the detectors ( < $3 \times 10^{-6}$ OD). The sampling electronics are similar to those described in reference 63.

A second probe beam (Figure 1), was derived from a 670 nm cw diode laser and used to compensate for pulse-to-pulse variations in the absorbed dose and for chemical dosimetry (see below). It propagated along the same optical path as the 800 nm probe beam. Both of these probe beams were used to measure TA signal from $e_{aq}^-$. However, whereas the 800 nm beam was used to obtain the TA on the picosecond time scale in a stroboscopic fashion, the 670 nm cw beam registered the same absorption on the nanosecond to microsecond time scales using a 30 MHz Si photodiode and a fast transient digitizer. TA signals as small as $5 \times 10^{-6}$ OD may be detected using this cw probe beam. The absolute measurement of the product yield (*G*-value) in pulse radiolysis is given in molecules per 100 eV of absorbed radiation (assuming linear energy transfer). To determine this quantity in pulse-probe experiments one has to know the deposition of the dose along the exact optical path of the probe beam. Since the energy/dose profile is complex (Figures 3 to 5) and depends on the sample, only chemical dosimetry is suitable. For a long-lived light-absorbing species like the hydrated electron, an option exists to observe the TA signal on a longer time scale using a cw laser beam. This signal can be used to determine electron absorbance at delay time > 500 ns which can be used to determine the absolute yield of hydrated electron at earlier times since the *G*-values for electron production at long delay times (ca. 2.65 by the end of the geminate stage, in the low-dose regime) are known from many previous experiments.[59]

Since pulse-to-pulse fluctuations in the beam charge are large (20-30% variance) and the repetition rate is relatively low (10 Hz), compensation of these fluctuations by normalization of the TA signal is important. Both short- and long- lived absorbance signals scale linearly with the dose. If both the pulsed and CW beams probe the same region of the sample, the two TA signals will track each other. This proportionality can be used to compensate for the variation in the dose deposition. Two approaches were explored simultaneously: (1) using the total charge of the electrons passing through the sample measured with the FC placed behind the sample cell, and (2) using cw-probed long-lived TA signal. The FC and long-lived TA signals were integrated using a boxcar integrator-averager or numerically, using the signal acquired on the oscilloscope. This detection system was tested by using the frequency doubled (400 nm) laser beam as a pump; a solution of octathiophen in toluene was used as a test system; the photodiode signal from the 400 nm light passing through the sample imitated the FC signal and a thermal lens signal in the sample imitated the long-lived TA signal. The two final Nd:YAG amplifiers were put in an unstable regime so that the pump power fluctuated by 100%. These tests demonstrated that with our detection setup the stabilization of the signal within 5% at 10 Hz acquisition could be readily achieved. This test also served to find the temporal overlap between the electron and 800 nm pulses. In the pulse radiolysis runs, we used the FC and TA signals not only to normalize the pulse-probe signal but also to reject 20-30% of electron pulses with low charge at the tail of the charge variation. All of these measures resulted in a considerable improvement of the signal-to-noise ratio. The typical error in the optical density measurement was $5 \times 10^{-4}$ for the average of 10 pulses, for a TA signal of $(5-20) \times 10^{-3}$ OD.

The general scheme for the acquisition of picosecond TA kinetics was as follows: the TW laser pulses were passed into the compressor in bunches of 10 at 5 Hz, and the TA signals and the charge on the FC recorded. Low-charge pulses were rejected, and the integrated TA signal for 670 nm light was used for normalization of the TA signal probed at 800 nm. The TW laser beam was then blocked for 10 shots to balance the reference and the probe dark signals. This sequence was repeated several times for each delay time of the probe beam. The quality of the TW laser beam mode was continually monitored for signs of optics damage at several pick-off points using CCD cameras. Most of the metal ejecta coming from the rotating copper disk blocking the laser was intercepted by a shield placed in front of the jet with a 5 mm diameter aperture for the 800 nm light, but some ejecta was eventually deposited on the parabolic mirror, limiting its lifetime to a few hours before replacement. A related problem was the slow deposition of graphitic carbon on the diffraction gratings during high-power operation. The latter was reversed by a periodic ultraviolet light and ozone treatment.

**V. RESULTS AND DISCUSSION**

In order to assess the time resolution experimentally achievable using our system, we have studied the decay kinetics of hydrated electrons ($e_{aq}^-$) generated in pulse radiolysis of liquid water and concentrated perchloric acid solutions at 25 °C. Upon the photoionization, the water molecule ejects the electron that subsequently localizes and thermalizes, yielding $e_{aq}^-$. The probe wavelength (800 nm) is close to the maximum (720 nm) of the absorption spectrum of the thermalized, fully hydrated species in room-temperature water. Other species generated in water radiolysis, such as OH and $OH_2$ radicals and H atoms do not absorb at 800 nm. Laser experiments in which water is biphotonically ionized using femtosecond pulses of ultraviolet light suggest that the local equilibrium between the cavity electron and water molecules around it is reached in < 1 ps, as judged from the evolution of the TA spectrum.[64] Before this equilibrium is fully reached, the pre-thermalized electron occupies a distorted, loose cavity and exhibits a broad TA spectrum that is shifted to the red from the final one.[64] As the electron thermalizes (with the characteristic time of ca. 200-300 fs) this TA spectrum continuously narrows and shifts to the blue; a similar shift can be observed for the thermalized electron when water is cooled from 100 to 0 °C.[44] At 800 nm, the evolution of the TA signal is complete in 1.5 ps.[64] The situation in pulse radiolysis is less clear, because the electrons are generated in spurs that thermalize with the water bulk on the picosecond time scale,[9] as the heat generated in the excitation and ionization events diffuses away. The tremendous amount of energy that is deposited within the spur may result in the overall thermalization rate that is slower than that observed in photolysis.

As discussed in the Introduction, in ultrafast pulse radiolysis (to a much greater degree than in ultrafast laser spectroscopy), the real time resolution depends on the sample, in particular, on the sample thickness. The deoxygenated sample was placed in a sealed fused silica cell (with 1 mm thick wall) that measured 10 mm in the direction of the electron beam and 2, 5, or 10 mm in the direction of the 800 nm probe light. The probe beam intersected the cell 200 μm away from the front window of the cell. Figure 6(a) exhibits the kinetics for $e_{aq}^-$ obtained in water for these three cells. The three kinetics shown in Figure 6(a) were obtained on different runs with slightly different electron pulse characteristics. We have checked that the TA signal is proportional to the sample thickness when the measurements were performed under identical conditions.

There are three factors that determine the rise time of the TA signal: (1) the formation time of the species and the time scale of their spectral evolution, (2) the pulse duration (and dispersion) of the electron beam at different points in the sample, and (3) the traveling time of the probe through the radiolytic zone. On the time scale of Figure 6(a), the formation of $e_{aq}^-$ can be considered as instantaneous (although that is not necessarily correct for the TA signal, as explained above). Assuming that the pulse duration for the electrons emerging from the jet is comparable in the duration to the TW laser pulse, and knowing the electron spectrum, the dispersion of the arrival times for the electrons at the sample was estimated as 3-4 ps (see section III). Given the large diameter of the electron beam at the sample (which is commensurate with the sample thickness) one would expect that the second factor determines the rise time of the TA signal observed. In principle, this rising part of the kinetics can be simulated from the calculated dose distribution (i.e., the distribution of $e_{aq}^-$) along the path of the probe beam. In reality, this distribution is not known exactly, and moreover it fluctuates from pulse to pulse. In order to make a comparison with other experimental systems the kinetics were fit by an error function which corresponds to a hypothetical experiment with a Gaussian pulse of electrons, infinitely thin sample and collinear beam geometry. The *$1/e^2$* Gaussian times obtained from the data in this fashion are 17.9, 8.3, and 4.4 ps for 10, 5, and 2 mm cells, respectively. Thus the "pulse width" scales with the sample thickness, supporting the assumption of short electron pulse duration. Thus the best "pulse width" obtained was on the order of 4 ps. In principle, this width can be shortened further by the use of thinner samples, better shaping of the electron spectrum, and collinear detection.

We have also performed radiolysis of 1 and 5 mol dm$^{-3}$ solutions of perchloric acid in a 10 mm optical path cell. Figure 6(b) shows the kinetics obtained. The fast decay is due to the reaction of $e_{aq}^-$ with the hydronium ($H_3O^+$) ions in the solution (this reaction yields H atoms that do not absorb at 800 nm).[65-67] The kinetics were fit to a single exponential convoluted with the error function. The obtained decay rates were 7.1x10$^9$ and $5.4 \times 10^{10}$ s$^{-1}$ in 1 and 5 mol dm$^{-3}$ solutions, respectively. These rate constants are in agreement with the previously published data obtained using a linac.[66,67] An intriguing feature of the 5 mol dm$^{-3}$ kinetics shown in Figure 6(b) is the apparent broadening of the "pulse width." This broadening has already been observed[67] with time resolution at least an order of magnitude slower than reported here. According to the analysis given therein, it cannot be accounted for by full consideration of the electron and probe beam propagation in the sample, possibly suggesting new chemistry. The presence of 5 mol dm$^{-3}$ of the anions considerably shifts the absorption spectrum of the electron to the blue, which results in greater sensitivity of the kinetics to the spectral evolution of $e_{aq}^-$ at 800 nm, as the spectrum shifts during the thermalization of $e_{aq}^-$ in the spur (and the spur itself) on the picosecond time scale. This interesting behavior will be pursued in our subsequent studies.

## VI. CONCLUSION

We have shown that a tabletop laser wakefield accelerator can achieve picosecond resolution for pulse-probe radiation chemistry experiments on condensed matter systems. Transient absorption kinetics were acquired in the stroboscopic fashion typical of other ultrafast methods, which required a robust design capable of continuous operation at 5-10 Hz. The time resolution was better or comparable to that for photocathode driven linacs; in fact, it is presently limited by the physics of electron deposition into the sample and the sensitivity of the detection rather than the source characteristics per se. If the sensitivity is improved, subpicosecond time resolution using this scheme will be possible. As it presently is, this time resolution (determined by the rise in the TA signal) is about 2 ps .

## VIII. ACKNOWLEDGEMENT.

We thank Dr. C. D. Jonah for the scientific motivation and constant reality checks, Prof. D. Umstadter for the use of his LWA to show that they can indeed be used for chemical measurements and Dr. V. Malka for their many valuable suggestions, Dr. L. Chen for the octathiophen solutions and Dr. S. Chemerisov, Dr. L Young, Dr. S. Southworth, A. Youngs and R. Lowers for their technical expertise and assistance. This

17.

**Figures.**

**Figure 1.**

The scheme of pulse radiolysis – Transient Absorption experiment using a tabletop laser wakefield accelerator (see sections II and IV for more detail).

**Figure 2.**

(a) The image of a LANEX radioluminescent screen placed behind a 400 μm thick aluminum shield at the position of the front end of the sample exposed to a single electron pulse. (b) Horizontal (solid) and Vertical (dashed) intensity slices from LANEX image.

**Figure 3.**

Energy spectra obtained from permanent magnet spectrometer for the electrons in (a) $2^o$ and (b) $13.5^o$ cones dispersed on the radiochromic film. Images are averaged over 100 shots. The oscillations are artifacts of the spectrum reconstruction.

**Figure 4.**

(a) The energy spectrum integrated over the entire electron beam reconstructed using the dosimetry method outlined in section III. (i) The spectrum for the electron beam incident on the copper disk. (ii) The spectrum of the electrons in the water sample at the position of the 800 nm probe beam. (b) The dispersion of arrival times for the electrons having the energy spectrum shown in Figure 3(a).

**Figure 5.**

(a) The rise of the transient absorbance ($\Delta OD$) from hydrated electron in pulse radiolysis of deoxygenated water in 10 mm (empty triangles), 5 mm (empty squares) and 2 mm (filled circles) optical path cells. See Section V for more detail. Each point represents the average of 100 pulses. The solid lines are least squares fits using error function. (b) The decay kinetics of observed in pulse radiolysis of 1 and 5 mol dm$^{-3}$ solutions of perchloric acid in water (filled squares and empty circles, respectively). The solid lines are exponential decay functions convoluted with a Gaussian. The electron rapidly decays in a reaction with hydronium ion.

# References.

Figure 1, Oulianov et al.

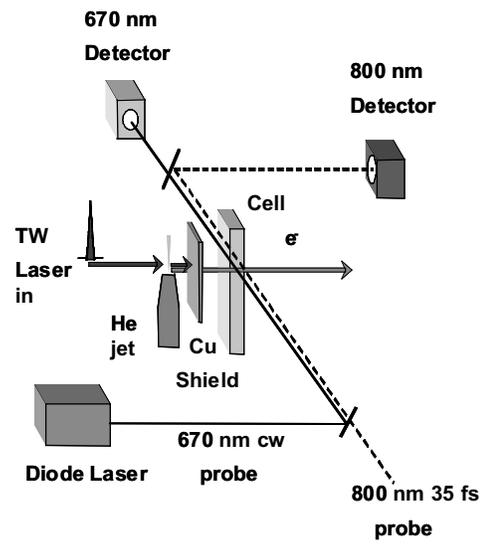

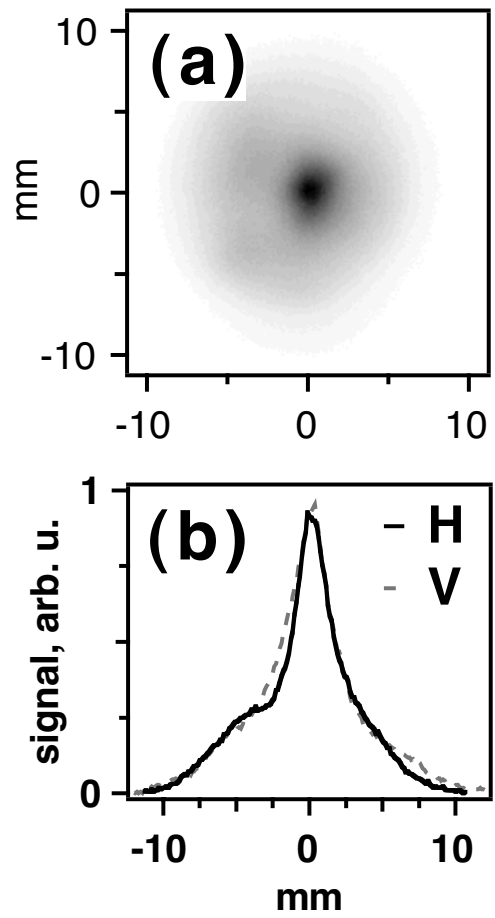

Figure 3, Oulianov et al.

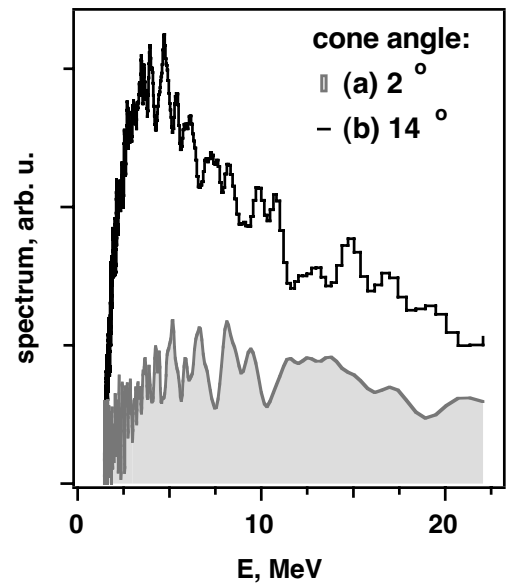

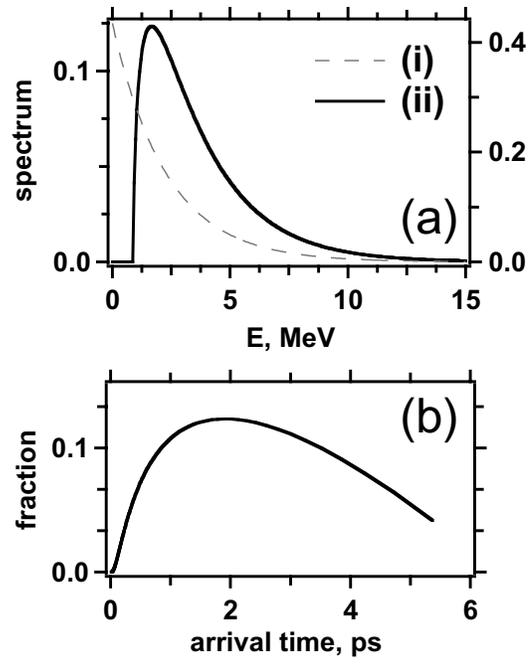

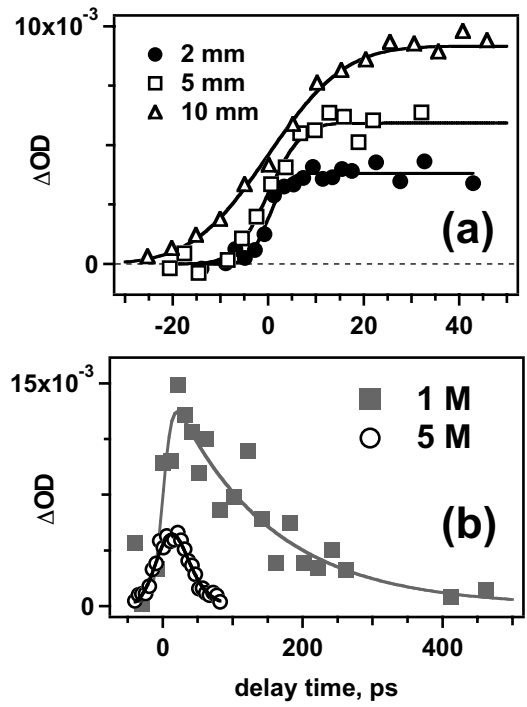